\documentstyle[aaspp4,12pt,psfig]{article}
\lefthead{Bower \& Backer}
\righthead{VSOP NRAO 530}
\begin{document}

\newcommand\degd{\ifmmode^{\circ}\!\!\!.\,\else$^{\circ}\!\!\!.\,$\fi}
\newcommand{\etal}{{\it et al.\ }}
\newcommand{\uv}{(u,v)}

\title{Space VLBI Observations Show $T_b > 10^{12} K$ in the Quasar NRAO~530}
\author{Geoffrey C. Bower}
\affil{Max Planck Institut f\"{u}r Radioastronomie, Auf dem H\"{u}gel 69, D 53121 Bonn Germany}
\author{Donald C. Backer}
\affil{Astronomy Department \& Radio Astronomy Laboratory, University of California, Berkeley, CA 94720}

\begin{abstract}
We present here space-based VLBI observations with VSOP and a southern
hemisphere ground array of the gamma-ray blazar NRAO 530 at 1.6 GHz and 5 GHz.
The brightness temperature of the core at 1.6 GHz is $5 \times 10^{11}$ K.
The size is near the minimum observable value in the
direction of NRAO~530 due to interstellar scattering.
The 5 GHz data show a single component with a brightness temperature of
$\sim 3 \times 10^{12}$ K, significantly in excess of the inverse Compton
limit and of the equipartition brightness temperature limit (Readhead 1994).
This is strong evidence for relativistic motion in a jet requiring
model-dependent Doppler boosting factors in the range 6 to 60.
We show that a simple homogeneous sphere probably does not model the
emission region accurately.  We favor instead an inhomogeneous jet model
with a Doppler boosting factor of 15.  
\end{abstract}


\section{Introduction}

Compact extragalactic radio sources typically exhibit brightness 
temperatures in the range of $10^{10}$ to $10^{11}$ K 
(Kellermann, Vermeulen, Zensus \& Cohen 1998).  These
brightness temperatures are near the limit of what can be measured
by ground-based very long baseline interferometry (VLBI) and 
are near to the limits imposed by intrinsic physical processes
(Readhead 1994).  Historically, the brightness temperature limit
has been attributed to the inverse Compton (IC) catastrophe, a process in which
the high energy electrons that produce the radio synchrotron photons
are rapidly cooled by scattering the same photons to higher energies
(e.g., Kellermann \& Pauliny-Toth 1969).  This process is considered
to be a likely source for the high energy gamma-rays
that are identified with very compact radio sources (von Montigny \etal 1995).
Brightness temperatures in excess of the IC limit have been 
interpreted as the effect of Doppler boosting in a relativistic
jet beamed towards the observer with a Doppler boosting factor $\delta \sim 10$.  

Readhead (1994) has shown that the actual distribution of
interferometrically-measured 
brightness temperatures does not correspond to the distribution
expected by a sample limited by IC scattering.  Instead,
the distribution indicates that brightness temperatures are 
governed by an unspecified mechanism that maintains equipartition
of energy between magnetic fields and particles in a synchrotron
emitting region which sets a brightness temperature limit that 
is $\sim 3$ to 10 times lower than the IC limit.  Brightness temperatures in
excess of this limit are again due to Doppler boosting.

Ground-based VLBI measurements of brightness temperatures have a 
limit of $\sim 5.6 \times 10^{10} (1+z) S$ K, where $S$ is
the flux density in Jy
and $z$ is the cosmological redshift.  
With techniques of super-resolution, lower limits to brightness
temperatures have been inferred that are greater than $10^{12}$ K
(e.g., Moellenbrock \etal 1996).
VLBI observations with an orbiting antenna permit the detection of
fully resolved components with brightness temperatures greater
than $10^{12}$ K (Linfield \etal 1989).  
Observations with 
the VLBI Space Observatory Programme (VSOP) have baselines greater than
two Earth diameters,
quadrupling the maximum detectable brightness temperature
(Hirabayashi 1998).  

We present here observations with VSOP of the blazar NRAO~530
at 1.6 GHz and 5 GHz.
NRAO~530 (J1733-1304) 
recently underwent a bright millimeter and radio wavelength
flare that appears to be correlated with the creation of 
a new component in the jet and with an increase in gamma-ray
activitity (Bower \etal 1997).
NRAO~530 is a $m_{pg} \approx 18.5$ mag QSO (Welch and Spinrad 1973)
with a redshift $z=0.902$ (Junkkarinen  1984).
It is
a gamma-ray active blazar with a bolometric luminosity of  
$0.9 \times 10^{48}h^{-2} {\rm\ ergs\ s^{-1}}$,
assuming isotropic emission (Nolan \etal  1996).
Throughout this paper we assume
$q_0=0.5$, $H_0 = 100 h {\rm \ km\ s^{-1}\ Mpc^{-1}}$ and $h=0.7$,
which implies an angular to linear scale conversion of
$6.0 {\rm \ pc \ mas^{-1}}$ and a luminosity distance of
4.4~Gpc.  

In Section 2, we present the observations and data reduction.
In Section 3, we discuss  the implications of the high brightness
temperature that we detect in NRAO 530.

\section{Observations and Data Reduction}

\subsection{Observations, Correlation and Fringe-Fitting}

NRAO~530 was observed by 
VSOP and an array of ground radio telescopes on 8 and 9 September 1997.  
Observations in left-circular polarization
at 1.6 GHz and 5 GHz were done in two separate
5 hour orbits of HALCA, the VSOP satellite.
The Usuda tracking station was used during the observations.
System temperatures were measured during adjacent tracking passes.
The ground radio telescopes (GRTs) consisted of Usuda (Japan), Seshan (China), 
Mopra and AT (Australia).  The ground-based
observations were approximately 8 hours in duration.
Figure~\ref{fig:uvcover} displays the $\uv$ coverage for the 5 GHz experiment.

The data were recorded in the S2 format with two intermediate
frequency (IF) bands each with a bandwidth of 16 MHz.
Data were correlated in March 1998 at Penticton (Carlson \etal 1998).  
The space-ground
and ground-ground baselines were accumulated with periods of 0.1 and 2 s,
respectively.  The data were binned in 128 and 256 frequency channels at 1.6 GHz
and 5 GHz, respectively.

Initial steps in data reduction were 
performed with the 15 April 1998 version of AIPS.  
Fringe fitting was done in two steps.  In the first step, 
singleband delays and rates
were determined for the GRTs alone.  In the
second step, the previous GRT solutions were applied and new
singleband delay and rate solutions for VSOP alone were found.
Fringe fitting simultaneously 
GRTs with VSOP produced bad solutions that did not correctly
eliminate  all single band delays.  

Strong fringes from VSOP to Usuda were consistently detected.  Solution
intervals of 60 s were employed at both frequencies.
At 1.6 GHz, fringe amplitudes to VSOP were 0.1 to 1 Jy.  At 5 GHz, fringe
amplitudes to VSOP were 1 to 3 Jy.  Residual fringe rates varied smoothly
between 0 and 90 mHz at 1.6 GHz and between -50 and 400 mHz at 5 GHz.
The maximum residual fringe rate corresponds to 
a coherence loss of $\sim 3$\% in an 0.1 s integration.  Measured
decorrelation over 60 s at the time of maximum residual rate on the
VSOP-Usuda baseline was 5.6\% at 5 GHz.
Residual fringe delays varied smoothly between 150 and 550 ns at 1.6 GHz
and between 0 and 400 ns at 5 GHz.

\subsection{Calibration, Imaging and Model-Fitting}

We show in Table~\ref{tab:amps} the system temperature and gain
information used for each antenna.  System temperatures for HALCA
were measured during tracking passes adjacent to the observations
and varied by less than 10\% over an entire
orbit.  Gain information for HALCA was taken from calibration
observations of Cygnus A during the period 21 October to 4 November  1997.

Data were averaged to 5 minutes before imaging and self-calibration.
Imaging and modeling were
performed with DIFMAP (Shepherd, Pearson \& Taylor 1994).  
The solutions converged quickly and fits to all baselines were good.
Amplitude scaling factors determined through self-calibration
are also included in Table~\ref{tab:amps}.  Only for the case of 
Mopra at 5 GHz were the corrections greater than 20\%.
We fit models to the self-calibrated visibility data.  
(Table~\ref{tab:comps}). 
We show in Figure~\ref{fig:radplot}
the visibility amplitude at 5 GHz as a function of $\uv$ distance along with
circular Gaussian models that bracket the best-fit model.

We have confidence in the 5 GHz amplitude self-calibration.
First, we note that fits to the 1.6 GHz and 5 GHz
data prior to amplitude self-calibration were consistent
with those for the self-calibrated data.  
Second, without
input of the zero-baseline flux the amplitude self-calibration
reduced the flux on the Mopra-AT baseline ($ b\approx 2 M\lambda$)
from $9.86 \pm 0.1$ to $7.48 \pm 0.06$ Jy.  
This latter value is consistent with a single-dish measurement
made in the same epoch (see below).  
The flux missing between this short baseline and the longer
baselines ($\sim 3.8$ Jy) must be contained in structures with
an angular size in the range 8 to 100 mas ($2 < b < 25 M\lambda$).
VLBA imaging at 8.4 GHz shows multiple structures with
$\sim 2$ Jy with a FWHM of $b\approx 25 M\lambda$ (Bower \etal 1998).  
If the flux and the size are proportional to wavelength,
then the missing flux is accounted for.

\subsection{Contemporaneous Spectrum}

The UMRAO reports observations of NRAO~530 within two weeks of the
VSOP observations (Aller \& Aller,
http://www.astro.lsa.umich.edu/obs/radiotel/umrao.html).  
They find fluxes of $7.34 \pm 0.13$, $8.49 \pm 0.11$,
and $7.49 \pm 0.06$ Jy at 4.8, 8.0 and 14.5 GHz.  Observations with
the BIMA millimeter interferometer give a flux of $\sim 5$ Jy at 86 GHz in
the same epoch.  We assume for the remainder of the paper
that the spectrum has a self-absorption turnover
at 8 GHz with a flux of 8.5 Jy and 
a high frequency spectral index
of $\alpha=-0.2$ for $S \propto \nu^{\alpha}$.

\section{Discussion}

\subsection{Comparison with Other Observations}

We tabulate the brightness temperatures of each of the components in
Table~\ref{tab:comps} using the expression 
\begin{equation}
T_b = 1.41 \times 10^9 {\rm K}
\left( 1 + z \right)
\left( S \over {\rm  Jy} \right) 
\left( \sigma_1 \sigma_2 \over {\rm mas^2} \right)^{-1}
\left( \lambda \over {\rm cm} \right)^2,
\end{equation}
where $z$ is the cosmological redshift of the source, $S$ is the
peak flux density, $\sigma_1$ and $\sigma_2$ are the FWHM sizes of
the component in major and minor axes and $\lambda$ is the observed
wavelength.  This and all subsequent brightness temperatures are given
in the rest frame of the host galaxy.

The 5 GHz brightness temperature is the highest ever measured
for NRAO~530 and is among the highest measured interferometrically.
Previous space VLBI observations of NRAO~530
with the TDRSS satellite at 15 GHz found a $T_b = 9 \times 10^{11}$ K
(Linfield \etal 1990).   
Previous 3 millimeter VLBI observations found a brightness
temperature 
of $4 \times 10^{11}$ K 
at the peak of the millimeter
flare (Bower \etal 1997).  
Contemporaneous measurements with the VLBA at 22 and 43 GHz give
brightness temperatures of $4 \times 10^{11}$ and $2 \times 10^{11}$ K, 
respectively (Bower \etal 1998). 
Given the limits to ground-based brightness temperatures, these
values also represent lower limits to the brightness temperature.  
These observations also indicate that the structure on sub-mas scales
has two components.  The 1.6 GHz and 5 GHz brightness temperatures are,
therefore, lower limits to the actual brightness temperature.

\subsection{Extrinsic Effects on the Brightness Temperature}

NRAO~530 is at a low galactic latitude in the direction of
the Galactic Center ($l=12^\circ, b=+11^\circ$) and, hence, may be
affected by significant interstellar scattering.  The expected scattering
size scales as $\lambda^{2.2}$ (Taylor \& Cordes 1993).  This is
marginally consistent with the power law index of the measured core sizes, 
$1.5 \pm 0.2$.
The expected scattering sizes are 1.8 mas at 1.6 GHz and 0.16 mas at
5 GHz.  Since the scattering sizes and 
the measured sizes are comparable and there is a large uncertainty
in the scattering sizes, we cannot reliably remove their
effects.  We conclude that the measured sizes at 1.6 GHz
and 5 GHz are near the minimum observable at this flux density
and that the brightness temperatures 
are lower limits to the intrinsic brightness temperature.

\subsection{Intrinsic Causes for Excess Brightness Temperature}

We can compare the observed  brightness temperatures to the limits 
imposed by intrinsic physical processes.  We consider first processes
associated with a homogenous sphere as discussed by Readhead (1994),
the IC catastrophe
and an unspecified mechanism that maintains equipartition of
energy.  Later, we will consider the limits imposed by
an inhomogeneous jet in energy equipartition.
The IC
catastrophe imposes a limit of $T_b \approx 5\times 10^{11} $K.
The equipartition requirement imposes a limit of 
$T_b \approx 0.5\times 10^{11}$ K.
These values are quite robust; they 
depend only weakly on the spectral parameters of the 
component.

The 5 GHz component significantly exceeds both of these limits.  
Doppler factors of 6 and 60 are necessary to accomodate the limits,
respectively.  We estimated previously through component proper motion
studies at millimeter wavelengths that $\beta_{app} \approx 7 h^{-1}$ 
(Bower \etal 1997) and have since refined this value through more
extensive monitoring to $\beta_{app} \approx 4 h^{-1}$  
(Bower \etal 1998).  We also estimated 
previously through synchrotron self-Compton arguments that 
$\delta > 11$ and through gamma-ray opacity arguments that
$\delta > 2.3$.  If we require that the apparent velocity is $4 h^{-1}c$,
we find $\theta=8\degd 9$ and $\gamma=4.4$ for $\delta=6$
and $\theta=0\degd 1$ and $\gamma=30$ for $\delta=60$; $\theta$
is the angle between the jet and the line of sight and $\gamma$
is the bulk Lorentz factor.

A Lorentz factor of 30 is at the limit of what can exist under
standard jet and accretion disk parameters.  Melia and K\"{o}nigl
(1989) calculate that thermal radiation from an accretion disk
produces a radiative drag on a jet that leads to terminal Lorentz
factors on the order of 10.  Further, objects with Lorentz factors
this large must be extremely rare since measured apparent velocities 
appear to have an upper limit of $\sim 10$ (Vermeulen \etal 1994).
We consider it more likely that the true Doppler boosting factor
is $\sim 6$ and that the limiting physical process in this
component is the IC catastrophe.

If this is true, then the component is
significantly far from the equipartition energy.  Following Readhead, we
calculate that the total energy exceeds the equipartition energy
value by a factor of $\sim 10^3$ and the magnetic field is less than
the equipartition value by a factor of $\sim 10^2$.  The particle
energy density exceeds the field energy density by a factor of $\sim 10^8$.
These conditions are far from what one expects under the typical
magnetized shock-in-jet scenario for the acceleration of relativistic 
electrons.

We now consider a third intrinsic explanation,
an inhomogeneous jet in energy equipartition.
In this case the brightness temperature limit is 
$3\times 10^{11} \delta^{5/6}$K (Blandford \& K\"{o}nigl 1979).
This limit is higher because of the special geometry
of the inhomogeneous jet which leads to a more peaked distribution
of the flux than in the case of the homogeneous sphere.  
The implied Doppler boosting
factor is then $\sim 15$ and $\theta \approx 2.5$ and $\gamma \approx 9$.
The expected size for such a jet at 5 GHz is $\sim 0.5$ mas,
in excellent agreement with our observations.  The implied
magnetic field strength at the radius of maximum brightness
temperature is 0.01 G and the total jet power is 
$7 \times 10^{45}$ erg s$^{-1}$.

We have securely measured a brightness temperature in excess of
$10^{12}$ K in the gamma-ray blazar NRAO~530
through observations with the VSOP orbiting radio telescope.
This is strong evidence for beamed relativistic motion in a jet.
We have considered several intrinsic causes for this extreme
brightness temperature.  We favor 
the inhomogeneous jet model of Blandford \& K\"{o}nigl which
produces a reasonable Doppler boosting
factor while maintaining energy equipartition.  
The equipartition
brightness temperature limit of Readhead does not apply in
this situation due to the extreme Doppler boosting factor
required.  Although this limit appears to apply in many other
sources, variability in NRAO~530 may allow departures from
this equipartition limit on timescales of a few years or more.
If the brightness temperature is limited by the IC
catastrophe in a homogeneous sphere, then the implied Doppler boosting factors
are more reasonable.  However, the departures from energy
equipartition are difficult to understand.
On the other hand, this scenario is favorable because it
provides a link between the millimeter and centimeter wavelength
variability and the gamma-ray activity observed in NRAO~530.  
One can speculate that blazars detected by EGRET are those in
which the equipartition brightness temperature limit is 
briefly superceded by the IC catastrophe limit.  
Space VLBI observations with VSOP and with future missions
such as ARISE (Ulvestad \& Linfield 1998) will be necessary to probe
the physical limits on high brightness temperature sources.

\acknowledgements 
This research has made use of data from the University of
Michigan Radio Astronomy Observatory which is supported by the National Science Foundation and by funds from the University of Michigan.
We gratefully acknowledge the VSOP Project, which is led by the Japanese
Institute of Space and Astronautical Science in cooperation with many
organizations and radio telescopes around the world.

\newpage

\figcaption[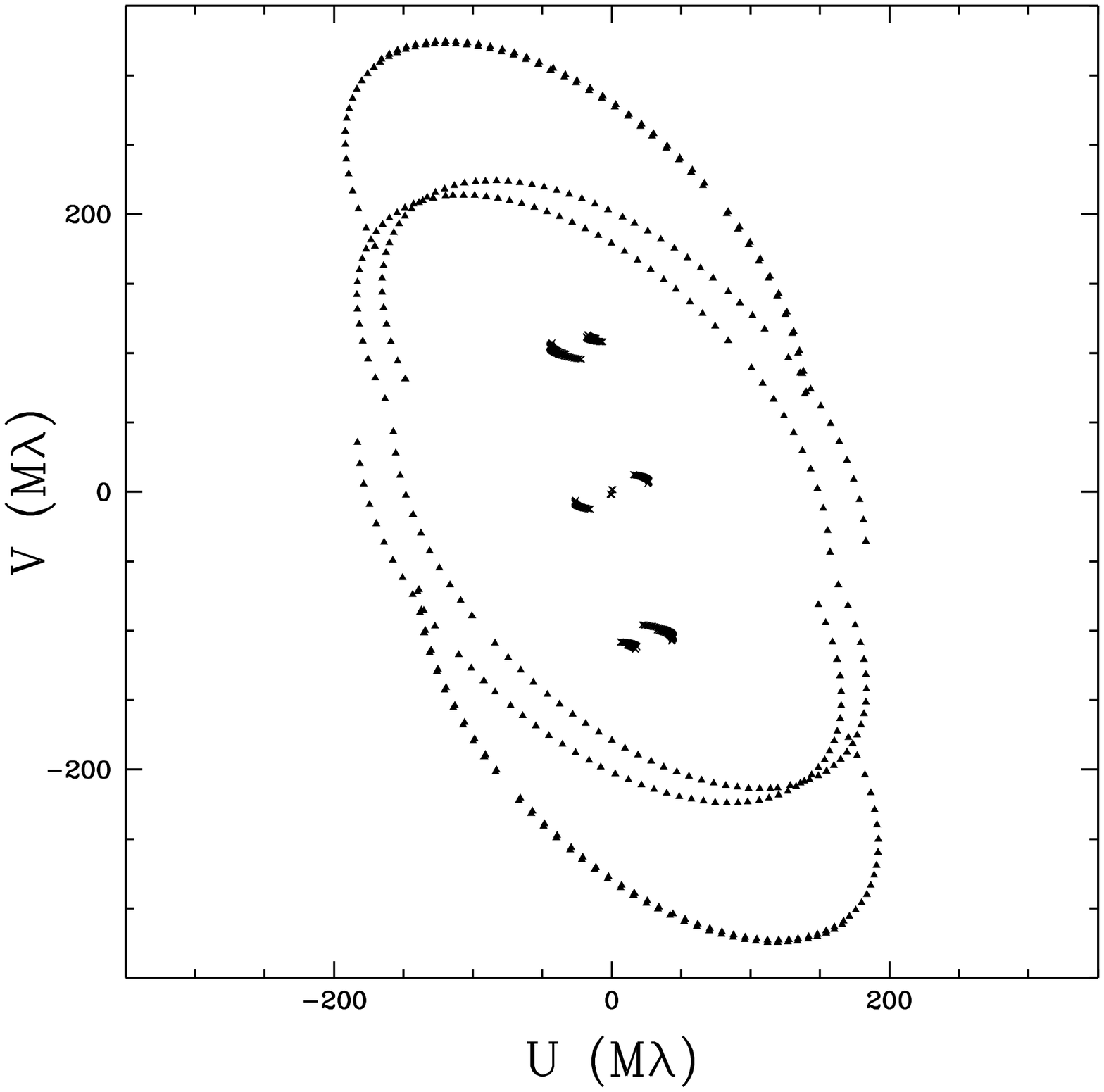]{The $\uv$ coverage at 5 GHz.  Triangles
indicate space-ground baselines and crosses indicate ground-ground
baselines.  Note that the VSOP-AT and VSOP-Mopra baselines overlap
and provide the greatest resolution.
\label{fig:uvcover}}

\figcaption[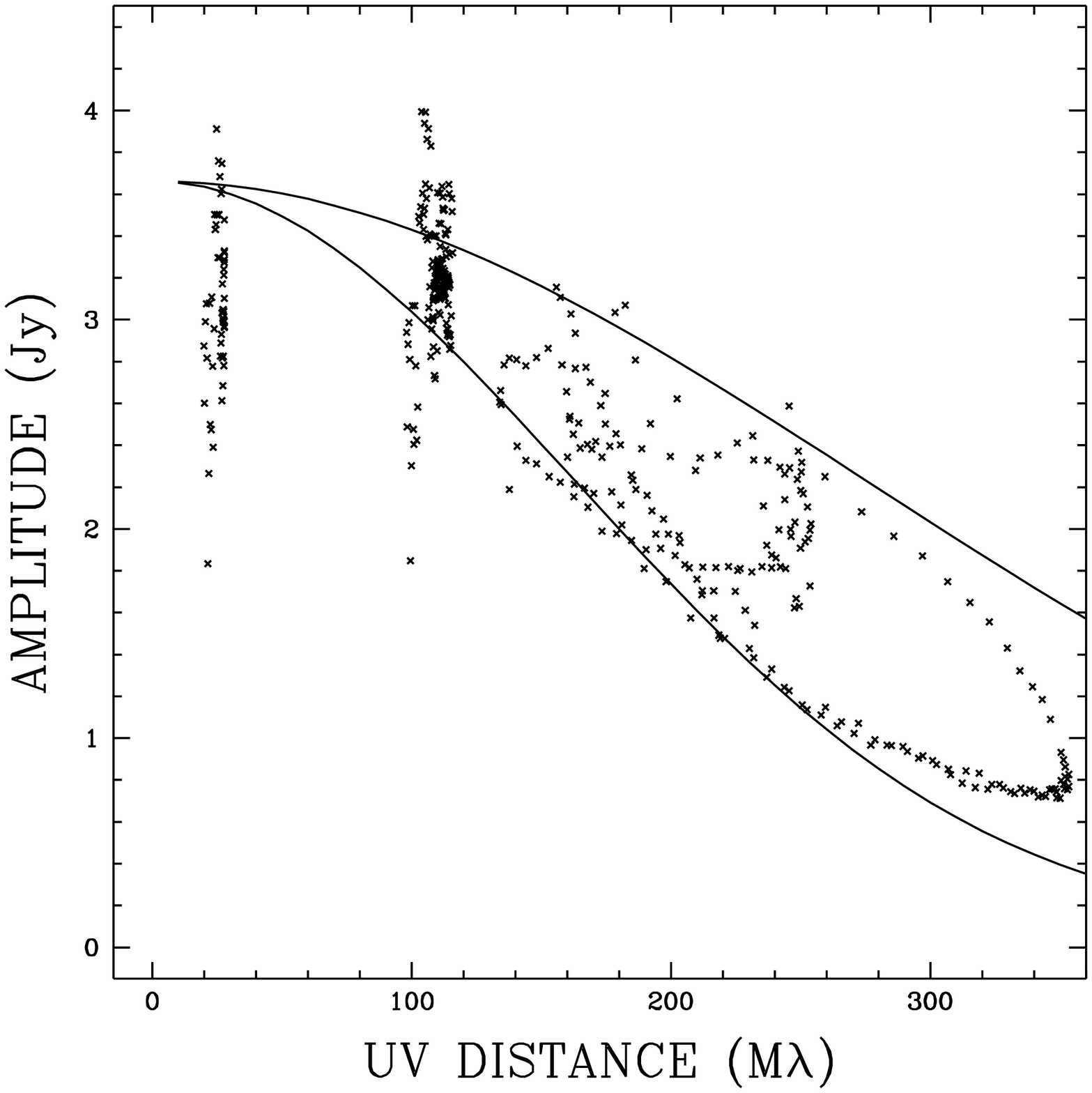]{Visibility amplitude as a function of $\uv$
distance at 5 GHz.  The solid lines are for a zero-baseline flux of 3.66 Jy
and circular Gaussian sizes of 0.47  and 0.28 mas, which correspond
to the major and minor axis FWHMs, respectively.  
\label{fig:radplot}}

\begin{deluxetable}{lrrllrr}
\tablecaption{Amplitude Calibration Parameters \label{tab:amps}}
\tablehead{
\colhead{Station} & \multicolumn{2}{c}{$T_{sys}$} & \multicolumn{2}{c}{Gain} &
\multicolumn{2}{c}{Scaling Factor}\\
 & \multicolumn{2}{c}{(K)} & \multicolumn{2}{c}{(K/Jy)} \\
& \colhead{1.6 GHz} & \colhead{5 GHz} & \colhead{1.6 GHz} & \colhead{5 GHz} 
& \colhead{1.6 GHz} & \colhead{5 GHz}\\
}
\startdata
HALCA & 82 & 104 & 0.0043 & 0.0062 & 0.97 & 1.16 \\
Seshan & 200 & 51 & 0.055 & 0.055 & 0.90 &  1.11\\
AT    &  40 & 45 & 0.55 & 0.55 & 0.92 & 1.06 \\
Mopra & 40 & 104 & 0.09 & 0.09 & 0.94 & 0.65 \\
Usuda & 81 & 109 & 0.63 & 0.63 & 1.13 & 0.91 \\
\enddata
\end{deluxetable}

\begin{deluxetable}{rrrrrrr}
\tablecaption{Model fits to components \label{tab:comps}}
\tablehead{
 \colhead{Flux} & \colhead{$r$} & \colhead{$\theta$} & 
\colhead{$\sigma_1$} & \colhead{$\sigma_2$} & \colhead{$\phi$} &
\colhead{$T_b$}\\
\colhead{(Jy)} & \colhead{(mas)} & \colhead{(deg)} &
\colhead{(mas)} & \colhead{(mas)} &  \colhead{(deg)} &
\colhead{($10^{11}$ K)} }
\startdata
\multicolumn{6}{c}{1.6 GHz} \\
\hline
   1.72 &    0.00  &  0.0 & 2.41 & 1.46 & 25.2 & 5 \\
  0.84  &    0.58  &  7.0 & 4.46 & 2.36 & 86.7 & 0.8 \\
  0.92  &    3.53  & 16.2 & 3.21 & 1.86 & 53.0 & 2 \\
\hline
\multicolumn{6}{c}{5 GHz} \\
\hline
     3.66 &  0.00  & 0.00 & 0.47 &  0.28 & 23.0 & 30 \\
\enddata
\tablecomments{The beam size at 1.6 GHz is $2.65 \times 0.98$ mas
in a position angle of $62^\circ$.  The beam size at 5 GHz is 
$0.78 \times 0.35$ mas in a position angle of $70^\circ$.}
\end{deluxetable}

\begin{figure}[p]
\mbox{\psfig{figure=uvcover.ps,width=\textwidth}}
\end{figure}

\begin{figure}[p]
\mbox{\psfig{figure=radplot.ps,width=\textwidth}}
\end{figure}


\begin{references}
\renewcommand{\etal}{{\it et al.}}

\reference{bland79} Blandford, R.D. \& K\"{o}nigl, A., 
1979, \apj,  232, 34.

\reference{bower97} Bower, G.C., Backer, D.C., Wright, M., Forster, J.R., 
Aller, H.D. \& Aller, M.F., 1997, ApJ, 484, 118.

\reference{bower98} Bower, G.C. \etal, 1998, in preparation.

\reference{carls98} Carlson, B., Dewdney, P.E., Perachenko, W.T., 
Burgess, T., Casorso, R., \& Cannon, W.H., 1998, in preparation.

\reference{hirab98} Hirabayashi, H., 1998, IAU 164, J.A. Zensus, G.B. Taylor, \& J.M. Wrobel, eds., ASP Conf., 144, 11.

\reference{junkk84} Junkkarinen, V., 1984, \pasp,  96, 539.

\reference{kelle69} Kellermann, K.I. \& Pauliny-Toth, I.I.K., 1969, \apj, 193,
43.

\reference{kelle98} Kellermann, K.I., Vermeulen, R.C., Zensus, J.A. \& 
Cohen, M.H., 1998, AJ, 115, 1295.

\reference{linfi89} Linfield, R.P. \etal, 1989, \apj, 336, 1105.

\reference{linfi90} Linfield, R.P. \etal, 1990, \apj, 358, 350.

\reference{marsc83} Marscher, A.P., 1983, \apj,  264, 296.

\reference{melia89} Melia, F., \& K\"{o}nigl, A., 1989, \apj, 340, 162.

\reference{moell96} Moellenbrock, G.A., et al., 1996, \aj, 111, 2174.

\reference{nolan96} Nolan, P.L. \etal, 1996, \apj,  459, 100.

\reference{readh94} Readhead, A.C.S., 1994, \apj, 426, 51.

\reference{sheph94} Shepherd, M.C., Pearson, T.J., \& Taylor, G.B., 1994, 
\baas, 26, 987.

\reference{taylo93} Taylor, J.H. \& Cordes, J.M., 1993, \apj, 411, 674.

\reference{ulves98} Ulvestad, J.S. \& Linfield, R.P., 1998, IAU 164, J.A. Zensus, G.B. Taylor, \& J.M. Wrobel, eds., ASP Conf., 144, 397.

\reference{verme94} Vermeulen, R.C., \& Cohen, M.H., 1994, \apj, 430, 467.

\reference{vonmo95a} von Montigny, C., \etal, 1995a, \apj, 440, 525.

\reference{welch73} Welch, W.J. and Spinrad, H., 1973, \pasp,  85, 456.

\end{references}
\end{document}